# Computing Small Unsatisfiable Cores in Satisfiability Modulo Theories

**Alessandro Cimatti**                                                                                     CIMATTI@FBK.EU
*FBK-IRST,*
*Via Sommarive 18, 38123 Povo, Trento, Italy*

**Alberto Griggio**                                                                                             GRIGGIO@FBK.EU
*FBK-IRST,*
*Via Sommarive 18, 38123 Povo, Trento, Italy*

**Roberto Sebastiani**                                                                                     RSEBA@DISI.UNITN.IT
*DISI, Università di Trento,*
*Via Sommarive 14, 38123 Povo, Trento, Italy*

## Abstract

The problem of finding small unsatisfiable cores for SAT formulas has recently received a lot of interest, mostly for its applications in formal verification. However, propositional logic is often not expressive enough for representing many interesting verification problems, which can be more naturally addressed in the framework of Satisfiability Modulo Theories, SMT. Surprisingly, the problem of finding unsatisfiable cores in SMT has received very little attention in the literature.

In this paper we present a novel approach to this problem, called the *Lemma-Lifting approach*. The main idea is to combine an SMT solver with an external propositional core extractor. The SMT solver produces the theory lemmas found during the search, dynamically lifting the suitable amount of theory information to the Boolean level. The core extractor is then called on the Boolean abstraction of the original SMT problem and of the theory lemmas. This results in an unsatisfiable core for the original SMT problem, once the remaining theory lemmas are removed.

The approach is conceptually interesting, and has several advantages in practice. In fact, it is extremely simple to implement and to update, and it can be interfaced with every propositional core extractor in a plug-and-play manner, so as to benefit for free of all unsat-core reduction techniques which have been or will be made available.

We have evaluated our algorithm with a very extensive empirical test on SMT-LIB benchmarks, which confirms the validity and potential of this approach.

## 1. Motivations and Goals

In the last decade we have witnessed an impressive advance in the efficiency of SAT techniques, which has brought large and previously-intractable problems at the reach of state-of-the-art SAT solvers. As a consequence, SAT solvers are now a fundamental tool in many industrial-strength applications, including most formal verification design flows for hardware systems, for equivalence, property checking, and ATPG. In particular, one of the most relevant problems in this context, thanks to its many important applications, is that of finding small *unsatisfiable cores*, that is, small unsatisfiable subsets of unsatisfiable sets of clauses.

Examples of such applications include use of SAT instead of BDDs for unbounded symbolic model checking (McMillan, 2002), automatic predicate discovery in abstraction refinement frameworks (McMillan & Amla, 2003; Wang, Kim, & Gupta, 2007), decision procedures (Bryant, Kroening, Ouaknine, Seshia, Strichman, & Brady, 2009), under-approximation and refinement in the context of bounded model checking of multi-threaded systems (Grumberg, Lerda, Strichman, & Theobald, 2005), debugging of design errors in circuits (Suelflow, Fey, Bloem, & Drechsler, 2008). For this reason, the problem of finding small unsat cores in SAT has been addressed by many authors in the recent years (Zhang & Malik, 2003; Goldberg & Novikov, 2003; Lynce & Marques-Silva, 2004; Oh, Mneimneh, Andraus, Sakallah, & Markov, 2004; Mneimneh, Lynce, Andraus, Marques-Silva, & Sakallah, 2005; Huang, 2005; Dershowitz, Hanna, & Nadel, 2006; Zhang, Li, & Shen, 2006; Biere, 2008; Gershman, Koifman, & Strichman, 2008; van Maaren & Wieringa, 2008; Asín, Nieuwenhuis, Oliveras, & Rodríguez Carbonell, 2008; Nadel, 2010).

The formalism of plain propositional logic, however, is often not suitable or expressive enough for representing many other real-world problems, including the verification of RTL designs, of real-time and hybrid control systems, and the analysis of proof obligations in software verification. Such problems are more naturally expressible as satisfiability problems in decidable first-order theories —Satisfiability Modulo Theories, SMT. Efficient SMT solvers have been developed in the last five years, called *lazy* SMT solvers, which combine a Conflict-Driven Clause Learning (CDCL) SAT solver based on the DPLL algorithm (Davis & Putnam, 1960; Davis, Logemann, & Loveland, 1962; Marques-Silva & Sakallah, 1996; Zhang & Malik, 2002) — hereafter simply "DPLL" — with ad-hoc decision procedures for many theories of interest (see, e.g., Nieuwenhuis, Oliveras, & Tinelli, 2006; Barrett & Tinelli, 2007; Bruttomesso, Cimatti, Franzén, Griggio, & Sebastiani, 2008; Dutertre & de Moura, 2006; de Moura & Bjørner, 2008).

Surprisingly, the problem of finding unsatisfiable cores in SMT has received virtually no attention in the literature. Although some SMT tools do compute unsat cores, this is done either as a byproduct of the more general task of producing proofs, or by modifying the embedded DPLL solver so that to apply basic propositional techniques to produce an unsat core. In particular, we are not aware of any work aiming at producing *small* unsatisfiable cores in SMT.

In this paper we present a novel approach addressing this problem, which we call the *Lemma-Lifting approach*. The main idea is to combine an SMT solver with an external propositional core extractor. The SMT solver stores and returns the theory lemmas it had to prove in order to refute the input formula; the external core extractor is then called on the Boolean abstraction of the original SMT problem and of the theory lemmas. Our algorithm is based on the following two key observations: i) the theory lemmas discovered by the SMT solver during search are *valid* clauses in the theory $\mathcal{T}$ under consideration, and therefore they do not affect the satisfiability of a formula in $\mathcal{T}$; and ii) the conjunction of the original SMT formula with all the theory lemmas is propositionally unsatisfiable. Therefore, the external (Boolean) core extractor finds an unsatisfiable core for (the Boolean abstraction of) the conjunction of the original formula and the theory lemmas, which can then be refined back into a subset of the original clauses by simply removing from it (the Boolean abstractions of) all theory lemmas. The result is an unsatisfiable core of the original SMT problem.

Although simple in principle, the approach is conceptually interesting: basically, the SMT solver is used to dynamically lift the suitable amount of theory information to the Boolean level. Furthermore, the approach has several advantages in practice: first, it is extremely simple to implement and to update; second, it is effective in finding small cores; third, the core extraction is not prone to complex SMT reasoning; finally, it can be interfaced with every propositional core extractor in a plug-and-play manner, so as to benefit for free of all unsat-core reduction techniques which have been or will be made available.

We have evaluated our approach by a very extensive empirical test on SMT-LIB benchmarks, in terms of both effectiveness (reduction in size of the cores) and efficiency (execution time). The results confirm the validity and versatility of this approach.

As a byproduct, we have also produced an extensive and insightful evaluation of the main Boolean unsat-core-generation tools currently available.

**Content.** The paper is organized as follows. In §2 and §3 we provide some background knowledge on techniques for SAT and SMT (§2), and for the extraction of unsatisfiable cores in SAT and in SMT (§3). In §4 we present and discuss our new approach and algorithm. In §5 we present and comment on the empirical tests. In §6 we conclude, suggesting some future developments.

## 2. SAT and SMT

Our setting is standard first order logic. A 0-ary function symbol is called a *constant*. A *term* is a first-order term built out of function symbols and variables. If $t_1, \ldots, t_n$ are terms and $p$ is a predicate symbol, then $p(t_1, \ldots, t_n)$ is an *atom*. A *formula* $\phi$ is built in the usual way out of the universal and existential quantifiers, Boolean connectives, and atoms. A *literal* is either an atom or its negation. We call a formula *quantifier-free* if it does not contain quantifiers, and *ground* if it does not contain free variables. A *clause* is a disjunction of literals. A formula is said to be in *conjunctive normal form* (CNF) if it is a conjunction of clauses. For every non-CNF formula $\varphi$, an equisatisfiable CNF formula $\psi$ can be generated in polynomial time (Tseitin, 1983).

We also assume the usual first-order notions of interpretation, satisfiability, validity, logical consequence, and theory, as given, e.g., by Enderton (1972). We write $\Gamma \models \phi$ to denote that the formula $\phi$ is a logical consequence of the (possibly infinite) set $\Gamma$ of formulas. A *first-order theory*, $\mathcal{T}$, is a set of first-order sentences. A structure $\mathcal{A}$ is a model of a theory $\mathcal{T}$ if $\mathcal{A}$ satisfies every sentence in $\mathcal{T}$. A formula is *satisfiable in $\mathcal{T}$* (or $\mathcal{T}$*-satisfiable*) if it is satisfiable in a model of $\mathcal{T}$. (We sometimes use the word "$\mathcal{T}$-formula" for a ground formula when we are interested in determining its $\mathcal{T}$-satisfiability.)

In what follows, with a little abuse of notation, we might sometimes denote conjunctions of literals $l_1 \wedge \ldots \wedge l_n$ as sets $\{l_1, \ldots, l_n\}$ and vice versa. If $\eta \equiv \{l_1, \ldots, l_n\}$, we might write $\neg \eta$ to mean $\neg l_1 \vee \ldots \vee \neg l_n$. Moreover, following the terminology of the SAT and SMT communities, we shall refer to predicates of arity zero as *propositional variables*, and to uninterpreted constants as *theory variables*.

Given a first-order theory $\mathcal{T}$ for which the (ground) satisfiability problem is decidable, we call a *theory solver for $\mathcal{T}$*, $\mathcal{T}$*-solver*, any tool able to decide the satisfiability in $\mathcal{T}$ of sets/conjunctions of ground atomic formulas and their negations — *theory literals* or $\mathcal{T}$*-literals* — in the language of $\mathcal{T}$. If the input set of $\mathcal{T}$-literals $\mu$ is $\mathcal{T}$-unsatisfiable, then a

```
1.      SatValue DPLL (formula φ, assignment μ) {
2.          while (1) {
3.              decide_next_branch(φ, μ);
4.              while (1) {
5.                  status = deduce(φ, μ);
6.                  if (status == sat)
7.                      return sat;
8.                  else if (status == conflict) {
9.                      ⟨blevel, η⟩ = analyze_conflict(φ, μ);
10.                     if (blevel < 0) return unsat;
11.                     else backtrack(blevel, φ, μ, η);
12.                 }
13.                 else break;
14.     }}}
```

Figure 1: Schema of a modern DPLL engine.

typical $\mathcal{T}$-solver not only returns unsat, but it also returns the subset $\eta$ of $\mathcal{T}$-literals in $\mu$ which was found $\mathcal{T}$-unsatisfiable. ($\eta$ is hereafter called a *theory conflict set*, and $\neg\eta$ a *theory conflict clause*.) If $\mu$ is $\mathcal{T}$-satisfiable, then $\mathcal{T}$-solver not only returns sat, but it may also be able to discover one (or more) deductions in the form $\{l_1, \ldots, l_n\} \models_\mathcal{T} l$, s.t. $\{l_1, \ldots, l_n\} \subseteq \mu$ and $l$ is an unassigned $\mathcal{T}$-literal. If so, we call $(\bigvee_{i=1}^{n} \neg l_i \vee l)$ a *theory-deduction clause*. Importantly, notice that both theory-conflict clauses and theory-deduction clauses are valid in $\mathcal{T}$. We call them *theory lemmas* or $\mathcal{T}$-*lemmas*.

*Satisfiability Modulo (the) Theory* $\mathcal{T}$ — $SMT(\mathcal{T})$ — is the problem of deciding the satisfiability of *Boolean combinations* of propositional atoms and theory atoms. Examples of useful theories are equality and uninterpreted functions ($\mathcal{EUF}$), difference logic ($\mathcal{DL}$) and linear arithmetic ($\mathcal{LA}$), either over the reals ($\mathcal{LA}(\mathbb{Q})$) or the integers ($\mathcal{LA}(\mathbb{Z})$), the theory of arrays ($\mathcal{AR}$), that of bit vectors ($\mathcal{BV}$), and their combinations. We call an $SMT(\mathcal{T})$ *tool* any tool able to decide $SMT(\mathcal{T})$. Notice that, unlike a $\mathcal{T}$-solver, an $SMT(\mathcal{T})$ tool must handle also Boolean connectives.

Hereafter we adopt the following terminology and notation. The symbols $\varphi, \psi$ denote $\mathcal{T}$-formulas, and $\mu, \eta$ denote sets of $\mathcal{T}$-literals; $\varphi^p, \psi^p$ denote propositional formulas, $\mu^p$, $\eta^p$ denote sets of propositional literals, which can be interpreted as truth assignments to variables.

## 2.1 Propositional Satisfiability with the DPLL Algorithm

Most state-of-the-art SAT procedures are evolutions of the Davis-Putnam-Longeman-Loveland (DPLL) procedure (Davis & Putnam, 1960; Davis et al., 1962). A high-level schema of a modern DPLL engine, adapted from the description given by Zhang and Malik (2002),

```
1.      SatValue Lazy_SMT_Solver (T-formula φ) {
2.          φᵖ = T2P(φ);
3.          while (DPLL(φᵖ, μᵖ) == sat) {
4.              ⟨ρ, η⟩ = T-solver(P2T(μᵖ))
5.              if (ρ == sat) then return sat;
6.              φᵖ = φᵖ ∧ T2P(¬η);
7.          };
8.          return unsat;
9.      };
```

Figure 2: A simplified schema for lazy $SMT(\mathcal{T})$ procedures.

is reported in Figure 1.[1] The Boolean formula $\varphi$ is in CNF; the assignment $\mu$ is initially empty, and it is updated in a stack-based manner.

In the main loop, `decide_next_branch`$(\varphi, \mu)$ chooses an unassigned literal $l$ from $\varphi$ according to some heuristic criterion, and adds it to $\mu$. (This operation is called *decision*, $l$ is called *decision literal* end the number of decision literals in $\mu$ after this operation is called the *decision level* of $l$.) In the inner loop, `deduce`$(\varphi, \mu)$ iteratively deduces literals $l$ deriving from the current assignment and updates $\mu$ accordingly; this step is repeated until either $\mu$ satisfies $\varphi$, or $\mu$ falsifies $\varphi$, or no more literals can be deduced, returning sat, conflict and unknown respectively. (The iterative application of Boolean deduction steps in deduce is also called *Boolean Constraint Propagation, BCP*.) In the first case, DPLL returns sat. In the second case, `analyze_conflict`$(\varphi, \mu)$ detects the subset $\eta$ of $\mu$ which caused the conflict (*conflict set*) and the decision level blevel to backtrack. If blevel $< 0$, then a conflict exists even without branching, and DPLL returns unsat. Otherwise, `backtrack`(blevel, $\varphi, \mu$) adds the clause $\neg\eta$ to $\varphi$ (*learning*) and backtracks up to blevel (*backjumping*), updating $\mu$ accordingly. (E.g., with the popular 1st-UIP schema, it backtracks to the smallest blevel where all but one literal in $\eta$ are assigned, and hence it deduces the negation of the remaining literal applying BCP on the learned clause $\neg\eta$; see Zhang, Madigan, Moskewicz, & Malik, 2001.) In the third case, DPLL exits the inner loop, looking for the next decision.

For a much deeper description of modern DPLL-based SAT solvers, we refer the reader, e.g., to the work of Zhang and Malik (2002).

### 2.2 Lazy Techniques for SMT

The idea underlying every lazy $SMT(\mathcal{T})$ procedure is that (a complete set of) the truth assignments for the propositional abstraction of $\varphi$ are enumerated and checked for satisfiability in $\mathcal{T}$; the procedure either returns sat if one $\mathcal{T}$-satisfiable truth assignment is found, or returns unsat otherwise.

We introduce the following notation. $\mathcal{T}2\mathcal{P}$ is a bijective function ("theory to propositional"), called *Boolean (or propositional) abstraction*, which maps propositional variables into themselves, ground $\mathcal{T}$-atoms into fresh propositional variables, and is homomorphic

---

1. We remark that many of the details provided here are not critical for understanding the rest of the paper, but are mentioned only for the sake of completeness.

w.r.t. Boolean operators and set inclusion. The function $\mathcal{P}2\mathcal{T}$ ("propositional to theory"), called *refinement*, is the inverse of $\mathcal{T}2\mathcal{P}$. (E.g., $\mathcal{T}2\mathcal{P}(\{((x - y \leq 3) \vee A_3), (A_2 \rightarrow (x = z))\}) = \{(B_1 \vee A_3), (A_2 \rightarrow B_2)\}$, $B_1$ and $B_2$ being fresh propositional variables, and $\mathcal{P}2\mathcal{T}(\{A_1, \neg A_2, \neg B_1, B_2\}) = \{A_1, \neg A_2, \neg(x - y \leq 3), (x = z)\}$.) In what follows, we shall use the "$^p$" superscript for denoting the Boolean abstraction of a formula/truth assignment (e.g., $\varphi^p$ denotes $\mathcal{T}2\mathcal{P}(\varphi)$, $\mu$ denotes $\mathcal{P}2\mathcal{T}(\mu^p)$). Given a $\mathcal{T}$-formula $\varphi$, we say that $\varphi$ is *propositionally unsatisfiable* when $\mathcal{T}2\mathcal{P}(\varphi) \models \bot$.

Figure 2 presents a simplified schema of a lazy $SMT(\mathcal{T})$ procedure, called the *off-line schema*. The propositional abstraction $\varphi^p$ of the input formula $\varphi$ is given as input to a SAT solver based on the DPLL algorithm (Davis et al., 1962; Zhang & Malik, 2002), which either decides that $\varphi^p$ is unsatisfiable, and hence $\varphi$ is $\mathcal{T}$-unsatisfiable, or returns a satisfying assignment $\mu^p$; in the latter case, $\mathcal{P}2\mathcal{T}(\mu^p)$ is given as input to $\mathcal{T}$-*solver*. If $\mathcal{P}2\mathcal{T}(\mu^p)$ is found $\mathcal{T}$-consistent, then $\varphi$ is $\mathcal{T}$-consistent. If not, $\mathcal{T}$-*solver* returns the conflict set $\eta$ which caused the $\mathcal{T}$-inconsistency of $\mathcal{P}2\mathcal{T}(\mu^p)$; the abstraction of the $\mathcal{T}$-lemma $\neg\eta$, $\mathcal{T}2\mathcal{P}(\neg\eta)$, is then added as a clause to $\varphi^p$. Then the DPLL solver is restarted from scratch on the resulting formula.

Practical implementations follow a more elaborated schema, called the *on-line schema* (see Barrett, Dill, & Stump, 2002; Audemard, Bertoli, Cimatti, Korniłowicz, & Sebastiani, 2002; Flanagan, Joshi, Ou, & Saxe, 2003). As before, $\varphi^p$ is given as input to a modified version of DPLL, and when a satisfying assignment $\mu^p$ is found, the refinement $\mu$ of $\mu^p$ is fed to the $\mathcal{T}$-*solver*; if $\mu$ is found $\mathcal{T}$-consistent, then $\varphi$ is $\mathcal{T}$-consistent; otherwise, $\mathcal{T}$-*solver* returns the conflict set $\eta$ which caused the $\mathcal{T}$-inconsistency of $\mathcal{P}2\mathcal{T}(\mu^p)$. Then the clause $\neg\eta^p$ is added in conjunction to $\varphi^p$, either temporarily or permanently ($\mathcal{T}$-*learning*), and, rather than starting DPLL from scratch, the algorithm backtracks up to the highest point in the search where one of the literals in $\neg\eta^p$ is unassigned ($\mathcal{T}$-*backjumping*), and therefore its value is (propositionally) implied by the others in $\neg\eta^p$.

An important variant of this schema (Nieuwenhuis et al., 2006) is that of building a "mixed Boolean+theory conflict clause", starting from $\neg\eta^p$ and applying the backward-traversal of the implication graph built by DPLL (Zhang et al., 2001), until one of the standard conditions (e.g., 1st UIP – Zhang et al., 2001) is achieved.

Other important optimizations are *early pruning* and *theory propagation*: the $\mathcal{T}$-solver is invoked also on (the refinement of) an intermediate assignment $\mu$: if it is found $\mathcal{T}$-unsatisfiable, then the procedure can backtrack, since no extension of $\mu$ can be $\mathcal{T}$-satisfiable; if not, and if the $\mathcal{T}$-*solver* performs a deduction $\{l_1, \ldots, l_n\} \models_\mathcal{T} l$ s.t. $\{l_1, \ldots, l_n\} \subseteq \mu$, then $\mathcal{T}2\mathcal{P}(l)$ can be unit-propagated, and the Boolean abstraction of the $\mathcal{T}$-lemma $(\bigvee_{i=1}^n \neg l_i \vee l)$ can be learned.

The on-line lazy $SMT(\mathcal{T})$ schema is a coarse description of the procedures underlying all the state-of-the-art lazy $SMT(\mathcal{T})$ tools like, e.g., BarceLogic, CVC3, MathSAT, Yices, Z3. The interested reader is pointed to, e.g., the work of Nieuwenhuis et al. (2006), Barrett and Tinelli (2007), Bruttomesso et al. (2008), Dutertre and de Moura (2006), and de Moura and Bjørner (2008), for details and further references, or to the work of Sebastiani (2007) and Barrett, Sebastiani, Seshia, and Tinelli (2009) for a survey.

## 3. Extracting Unsatisfiable Cores

Without loss of generality, in the following we consider only formulas in CNF. Given an unsatisfiable CNF formula $\varphi$, we say that an unsatisfiable CNF formula $\psi$ is an *unsatisfiable core* of $\varphi$ iff $\varphi = \psi \wedge \psi'$ for some (possibly empty) CNF formula $\psi'$. Intuitively, $\psi$ is a subset of the clauses in $\varphi$ causing the unsatisfiability of $\varphi$. An unsatisfiable core $\psi$ is *minimal* iff the formula obtained by removing any of the clauses of $\psi$ is satisfiable. A *minimum* unsat core is a minimal unsat core with the smallest possible cardinality.

### 3.1 Techniques for Unsatisfiable-Core Extraction in SAT

In the last few years, several algorithms for computing small, minimal or minimum unsatisfiable cores of propositional formulas have been proposed. In the approach of Zhang and Malik (2003) and Goldberg and Novikov (2003), they are computed as a byproduct of a DPLL-based proof-generation procedure. The computed unsat core is simply the collection of all the original clauses that the DPLL solver used to derive the empty clause by resolution. The returned core is not minimal in general, but it can be reduced by iterating the algorithm until a fixpoint, using as input of each iteration the core computed at the previous one. The algorithm of Gershman et al. (2008), instead, manipulates the resolution proof so as to shrink the size of the core, using also a fixpoint iteration as Zhang and Malik (2003) to further enhance the quality of the results. Oh et al. (2004) present an algorithm to compute *minimal* unsat cores. The technique is based on modifications of a standard DPLL engine, and works by adding some extra variables (selectors) to the original clauses, and then performing a branch-and-bound algorithm on the modified formula. The procedure presented by Huang (2005) extracts minimal cores using BDD manipulation techniques, removing one clause at a time until the remaining core is minimal. The construction of a minimal core by Dershowitz et al. (2006) also uses resolution proofs, and it works by iteratively removing from the proof one input clause at a time, until it is no longer possible to prove inconsistency. When a clause is removed, the resolution proof is modified to prevent future use of that clause.

As far as the the computation of *minimum* unsatisfiable cores is concerned, the algorithm of Lynce and Marques-Silva (2004) searches all the unsat cores of the input problem; this is done by introducing selector variables for the original clauses, and by increasing the search space of the DPLL solver to include also such variables; then, (one of) the unsatisfiable subformulas with the smallest number of selectors assigned to true is returned. The approach described by Mneimneh et al. (2005) instead is based on a branch-and-bound algorithm that exploits the relation between maximal satisfiability and minimum unsatisfiability. The same relation is used also by the procedure of Zhang et al. (2006), which is instead based on a genetic algorithm.

### 3.2 Techniques for Unsatisfiable-Core Extraction in SMT

To the best of our knowledge, there is no literature explicitly addressing the problem of computing unsatisfiable cores in SMT [2]. However, four SMT solvers (i.e. CVC3, Barrett & Tinelli, 2007, MathSAT, Bruttomesso et al., 2008, Yices, Dutertre & de Moura, 2006 and Z3, de Moura & Bjørner, 2008) support unsat core generation[3]. In the following, we describe the underlying approaches, that generalize techniques for propositional UC extraction. We preliminarily remark that none of these solvers aims at producing minimal or minimum unsat cores, nor does anything to reduce their size.

---

2. Except for a previous short version of the present paper (Cimatti, Griggio, & Sebastiani, 2007).

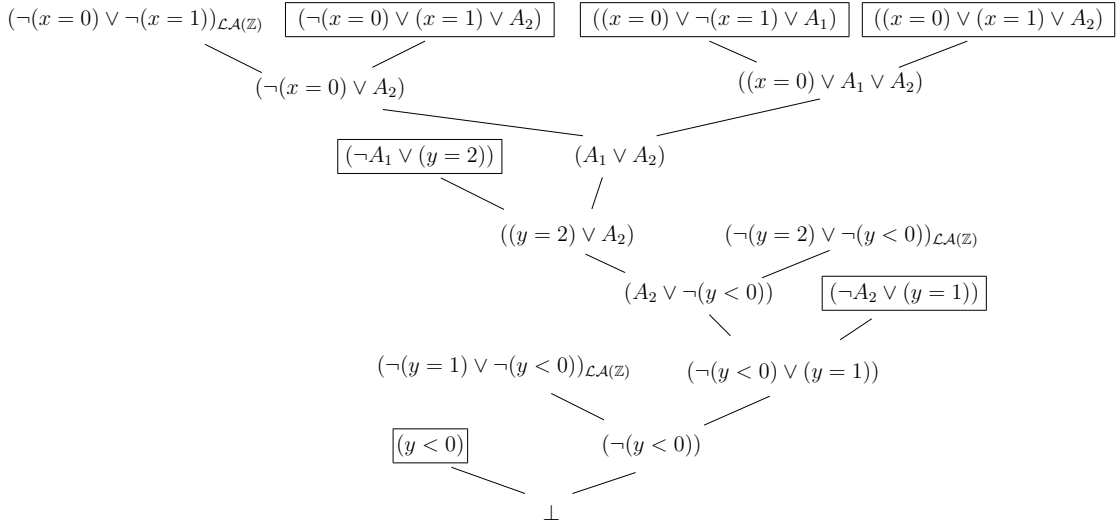

Figure 3: Resolution proof for the SMT formula (1) found by MathSAT. Boxed clauses correspond to the unsatisfiable core.

Strictly related with this work, Liffiton and Sakallah (2008) presented a general technique for enumerating all minimal unsatisfiable subsets of a given inconsistent set of constraints, which they implemented in the tool CAMUS. Although the description of the properties and algorithms focuses on pure SAT, the authors remark that the approach extends easily to SMT, and that they have implemented inside CAMUS a SMT version of the procedure. Therefore in the following we briefly describe also their approach.

### 3.2.1 Proof-Based UC Extraction.

CVC3 and MathSAT can run in proof-producing mode, and compute unsatisfiable cores as a byproduct of the generation of proofs. Similarly to the approach of Zhang and Malik (2003), the idea is to analyze the proof of unsatisfiability backwards, and to return an unsatisfiable core that is a collection of the assumptions (i.e. the clauses of the original problem) that are used in the proof to derive contradiction.

---

3. The information reported here on the computation of unsat cores in CVC3, Yices and Z3 comes from private communications from the authors and from the user manual of CVC3.

**Example 1** *In order to show how the described approaches work, consider this small unsatisfiable $SMT(\mathcal{T})$ formula, where $\mathcal{T}$ is $\mathcal{LA}(\mathbb{Z})$:*

$$((x=0) \vee \neg(x=1) \vee A_1) \wedge ((x=0) \vee (x=1) \vee A_2) \wedge (\neg(x=0) \vee (x=1) \vee A_2) \wedge$$
$$(\neg A_2 \vee (y=1)) \wedge (\neg A_1 \vee (x+y>3)) \wedge (y<0) \wedge (A_2 \vee (x-y=4)) \wedge$$
$$((y=2) \vee \neg A_1) \wedge (x \geq 0), \quad (1)$$

*where $x$ and $y$ are real variables and $A_1$ and $A_2$ are Booleans.*

*In the proof-based approach, a resolution proof of unsatisfiability is built during the search. E.g., Figure 3 shows the proof tree found by* MathSAT. *The leaves of the tree are either original clauses (boxed in the Figure) or $\mathcal{LA}(\mathbb{Z})$-lemmas (denoted with the $\mathcal{LA}(\mathbb{Z})$ suffix). The unsatisfiable core is built by collecting all the original clauses appearing as leaves in the proof. In this case, this is:*

$$\{((x=0) \vee \neg(x=1) \vee A_1), ((x=0) \vee (x=1) \vee A_2), (\neg(x=0) \vee (x=1) \vee A_2),$$
$$(\neg A_2 \vee (y=1)), (y<0), ((y=2) \vee \neg A_1)\}. \quad (2)$$

*In this case, the unsat core is minimal.*

### 3.2.2 Assumption-Based UC Extraction

The approach used by Yices (Dutertre & de Moura, 2006) and Z3 (de Moura & Bjørner, 2008) is an adaptation of the method by Lynce and Marques-Silva (2004): for each clause $C_i$ in the problem, a new Boolean "selector" variable $S_i$ is created; then, each $C_i$ is replaced by $(S_i \rightarrow C_i)$; finally, before starting the search each $S_i$ is forced to true. In this way, when a conflict at decision level zero is found by the DPLL solver the conflict clause contains only selector variables, and the unsat core returned is the union of the clauses whose selectors appear in such conflict clause.

**Example 2** *Consider again the formula (1) of Example 1. In the assumption-based approach, each of the 9 input clauses is augmented with an extra variable $S_i$, which is asserted to true at the beginning of the search. The formula therefore becomes:*

$$\bigwedge_i S_i \wedge$$
$$(S_1 \rightarrow ((x=0) \vee \neg(x=1) \vee A_1)) \wedge (S_2 \rightarrow ((x=0) \vee (x=1) \vee A_2)) \wedge$$
$$(S_3 \rightarrow (\neg(x=0) \vee (x=1) \vee A_2)) \wedge (S_4 \rightarrow (\neg A_2 \vee (y=1))) \wedge \quad (3)$$
$$(S_5 \rightarrow (\neg A_1 \vee (x+y>3))) \wedge (S_6 \rightarrow (y<0)) \wedge$$
$$(S_7 \rightarrow (A_2 \vee (x-y=4))) \wedge (S_8 \rightarrow ((y=2) \vee \neg A_1)) \wedge (S_9 \rightarrow (x \geq 0))$$

*The final conflict clause generated by conflict analysis (Zhang et al., 2001) is:* [4]

$$\neg S_1 \vee \neg S_2 \vee \neg S_3 \vee \neg S_4 \vee \neg S_6 \vee \neg S_7 \vee \neg S_8, \quad (4)$$

---
4. using Yices.

*corresponding to the following unsat core:*

$$\{((x=0) \vee \neg(x=1) \vee A_1), ((x=0) \vee (x=1) \vee A_2), (\neg(x=0) \vee (x=1) \vee A_2),$$
$$(\neg A_2 \vee (y=1)), (y<0), (A_2 \vee (x-y=4)), ((y=2) \vee \neg A_1)\}. \quad (5)$$

*Notice that this is not minimal, because of the presence of the redundant clause $(A_2 \vee (x-y=4))$, corresponding to $\neg S_7$ in the final conflict clause (4).*

**Remark 1** *The idea behind the two techniques just illustrated is essentially the same. Both exploit the implication graph built by DPLL during conflict analysis to detect the subset of the input clauses that were used to decide unsatisfiability. The main difference is that in the proof-based approach this is done by explicitly constructing the proof tree, while in the activation-based one this can be done "implicitly" by "labeling" each of the original clauses. For a deeper comparison between these two approaches (and some variants of them), we refer the reader to the work of Asín et al. (2008) and Nadel (2010).*

### 3.2.3 The CAMUS Approach for Extracting All Minimal UC's.

A completely different approach, aiming at generating all minimal UC's of some given inconsistent set of propositional clauses $\Phi$, is presented by Liffiton and Sakallah (2008) and implemented in the tool CAMUS. In a nutshell, the approach works in two distinct phases:

(a) enumerate the set $M$ of all *Minimal Correction Subsets (MCS's)* of $\Phi$. [5] This is performed by a specialized algorithm, using as backend engine an incremental SAT solver able to handle also AtMost constraints;

(b) enumerate the set $U$ of all the minimal UC's of $\Phi$ as *minimal hitting sets* of the set $M$. This is also performed by a specialized algorithm. Alternatively, another algorithm can produce from $M$ only one minimal UC with much less effort.

It is important to notice that both sets $M$ and $U$ returned can be exponentially big wrt. the size of $\Phi$. Thus, the procedure may produce an exponential amount of MCS's during phase (a) before producing one UC. To this extent, the authors provide also some modified and more efficient version of the technique, which sacrifice the completeness of the approach. We refer the reader to the work of Liffiton and Sakallah (2008) for a more detailed explanation of this technique and of its features.

As mentioned above, although the description of the algorithms focuses on pure SAT, the authors remark that the approach extends easily to SMT, and that they have implemented inside CAMUS a version of the algorithm working also for SMT, using Yices as backend SMT solver. Unfortunately, they provide no details of such an extension. [6]

---

5. A MCS $\Psi$ of an unsatisfiable set of constraint $\Phi$ is the complement set of a maximal consistent subset of $\Phi$: $\Phi \setminus \Psi$ is consistent and, for every $C_i \in \Psi$, $\Phi \setminus (\Psi \setminus C_i)$ is inconsistent (Liffiton & Sakallah, 2008).
6. See §10 "Conclusions and Future Work." of the article by Liffiton and Sakallah (2008).

**Example 3** *Consider again the $\mathcal{LA}(\mathbb{Z})$-formula (1) of Example 1 in form of clause set*

$$\Phi \stackrel{def}{=} \left\{ \begin{array}{ll} c_1: & (x=0) \vee \neg(x=1) \vee A_1, \\ c_2: & (x=0) \vee (x=1) \vee A_2, \\ c_3: & \neg(x=0) \vee (x=1) \vee A_2, \\ c_4: & \neg A_2 \vee (y=1), \\ c_5: & \neg A_1 \vee (x+y>3), \\ c_6: & (y<0), \\ c_7: & A_2 \vee (x-y=4), \\ c_8: & (y=2) \vee \neg A_1, \\ c_9: & (x \geq 0) \end{array} \right\}. \quad (6)$$

*When run on* (6), CAMUS *returns the following two minimal UC's:*

$$uc_1 \stackrel{def}{=} \left\{ \begin{array}{ll} c_1: & (x=0) \vee \neg(x=1) \vee A_1, \\ c_2: & (x=0) \vee (x=1) \vee A_2, \\ c_3: & \neg(x=0) \vee (x=1) \vee A_2, \\ c_4: & \neg A_2 \vee (y=1), \\ c_5: & \neg A_1 \vee (x+y>3), \\ c_6: & (y<0) \end{array} \right\}, \quad uc_2 \stackrel{def}{=} \left\{ \begin{array}{ll} c_1: & (x=0) \vee \neg(x=1) \vee A_1, \\ c_2: & (x=0) \vee (x=1) \vee A_2, \\ c_3: & \neg(x=0) \vee (x=1) \vee A_2, \\ c_4: & \neg A_2 \vee (y=1), \\ c_6: & (y<0), \\ c_8: & (y=2) \vee \neg A_1 \end{array} \right\}. \quad (7)$$

*(Notice that $uc_2$ is identical to the UC found in Example 1.)*
*We understand from Liffiton and Sakallah (2008) that, in order to produce $uc_1$ and $uc_2$, CAMUS enumerates first (not necessarily in this order) the following set of MCS's:*

$$\{\{c_1\}, \{c_2\}, \{c_3\}, \{c_4\}, \{c_6\}, \{c_5, c_8\}\} \quad (8)$$

*and then computes $uc_1$ and $uc_2$ as minimal hitting sets of* (8).
*Notice that* (8) *is a set of MCS's because $\Phi$, $\Phi \setminus \{c_5\}$ and $\Phi \setminus \{c_8\}$ are $\mathcal{LA}(\mathbb{Z})$-inconsistent, and*

$$\begin{array}{ll} \{A_1 = \bot, A_2 = \bot, x=1, y=-3\} & \models_{\mathcal{LA}(\mathbb{Z})} \Phi \setminus \{c_1\}, \\ \{A_1 = \bot, A_2 = \bot, x=2, y=-6\} & \models_{\mathcal{LA}(\mathbb{Z})} \Phi \setminus \{c_2\}, \\ \{A_1 = \bot, A_2 = \bot, x=0, y=-4\} & \models_{\mathcal{LA}(\mathbb{Z})} \Phi \setminus \{c_3\}, \\ \{A_1 = \bot, A_2 = \top, x=0, y=-1\} & \models_{\mathcal{LA}(\mathbb{Z})} \Phi \setminus \{c_4\}, \\ \{A_1 = \bot, A_2 = \top, x=3, y=1\} & \models_{\mathcal{LA}(\mathbb{Z})} \Phi \setminus \{c_6\}, \\ \{A_1 = \top, A_2 = \bot, x=1, y=-1\} & \models_{\mathcal{LA}(\mathbb{Z})} \Phi \setminus \{c_5, c_8\}. \end{array}$$

*Moreover, it contains all MCS's of $\Phi$ because also $\Phi \setminus \{c_9\}$, $\Phi \setminus \{c_5, c_9\}$ and $\Phi \setminus \{c_8, c_9\}$ are $\mathcal{LA}(\mathbb{Z})$-inconsistent.*

## 4. A Novel Approach to Building Unsatisfiable Cores in SMT

We present a novel approach, called the *Lemma-Lifting approach*, in which the unsatisfiable core is computed *a posteriori* w.r.t. the execution of the SMT solver, and only if the formula has been found $\mathcal{T}$-unsatisfiable. This is done by means of an external (and possibly optimized) propositional unsat core extractor.

### 4.1 The Main Ideas

In the following, we assume that a lazy $SMT(\mathcal{T})$ procedure has been run over a $\mathcal{T}$-unsatisfiable set of $SMT(\mathcal{T})$ clauses $\varphi =_{def} \{C_1, \ldots, C_n\}$, and that $D_1, \ldots, D_k$ denote all the $\mathcal{T}$-lemmas, both theory-conflict and theory-deduction clauses, which have been returned by the $\mathcal{T}$-solver during the run. (Notice that, by definition, $\mathcal{T}$-lemmas are $\mathcal{T}$-valid clauses.) In case of mixed Boolean+theory-conflict clauses (Nieuwenhuis et al., 2006) (see § 2.2), the $\mathcal{T}$-lemmas are those returned by the $\mathcal{T}$-solver that have been used to compute the mixed Boolean+theory-conflict clause, including the initial theory-conflict clause and the theory-deduction clauses corresponding to the theory-propagation steps performed.[7]
Under the above assumptions, two simple facts hold.

(i) Since the $\mathcal{T}$-lemmas $D_i$ are valid in $\mathcal{T}$, they do not affect the $\mathcal{T}$-satisfiability of a formula: $(\psi \wedge D_i) \models_\mathcal{T} \bot \iff \psi \models_\mathcal{T} \bot$.
(ii) The conjunction of $\varphi$ with all the $\mathcal{T}$-lemmas $D_1, \ldots, D_k$ is propositionally unsatisfiable: $\mathcal{T}2\mathcal{P}(\varphi \wedge \bigwedge_{i=1}^n D_i) \models \bot$.

Fact (i) is self-evident. Fact (ii) is the termination condition of all lazy SMT tools when the input formula is $\mathcal{T}$-unsatisfiable. In the off-line schema of Figure 2, the procedure ends when DPLL establishes that $\mathcal{T}2\mathcal{P}(\varphi \wedge \bigwedge_{i=1}^n D_i)$ is unsatisfiable, each $D_i$ being the negation of the theory-conflict set $\eta_i$ returned by the $i$-th call to the $\mathcal{T}$-solver. Fact (ii) generalizes to the on-line schema, noticing that $\mathcal{T}$-backjumping on a theory-conflict clause $D_i$ produces an analogous effect as re-invoking DPLL on $\varphi^p \wedge \mathcal{T}2\mathcal{P}(D_i)$, whilst theory propagation on a deduction $\{l_1, \ldots, l_k\} \models_\mathcal{T} l$ can be seen as a form on unit propagation on the theory-deduction clause $\mathcal{T}2\mathcal{P}(\bigvee_i \neg l_i \vee l)$.

**Example 4** *Consider again formula (1) of Example 1. In order to decide its unsatisfiability, MATHSAT generates the following set of $\mathcal{LA}(\mathbb{Z})$-lemmas:*

$$\{(\neg(x = 1) \vee \neg(x = 0)), (\neg(y = 2) \vee \neg(y < 0)), (\neg(y = 1) \vee \neg(y < 0))\}. \quad (9)$$

*Notice that they are all $\mathcal{LA}(\mathbb{Z})$-valid (fact (i)). Then, the Boolean abstraction of (1) is conjoined with the Boolean abstraction of these $\mathcal{LA}(\mathbb{Z})$-lemmas, resulting in the following propositional formula:*

$$(B_1 \vee \neg B_2 \vee A_1) \wedge (B_1 \vee B_2 \vee A_2) \wedge (\neg B_1 \vee B_2 \vee A_2) \wedge (\neg A_2 \vee B_3) \wedge$$
$$(\neg A_1 \vee B_4) \wedge B_5 \wedge (A_2 \vee B_6) \wedge (B_7 \vee \neg A_1) \wedge B_8 \wedge$$
$$(\neg B_2 \vee \neg B_1) \wedge (\neg B_7 \vee \neg B_5) \wedge (\neg B_3 \vee \neg B_5), \quad (10)$$

*where:*

$$B_1 \stackrel{def}{=} \mathcal{T}2\mathcal{P}(x = 0) \qquad B_5 \stackrel{def}{=} \mathcal{T}2\mathcal{P}(y < 0)$$
$$B_2 \stackrel{def}{=} \mathcal{T}2\mathcal{P}(x = 1) \qquad B_6 \stackrel{def}{=} \mathcal{T}2\mathcal{P}(x - y = 4)$$
$$B_3 \stackrel{def}{=} \mathcal{T}2\mathcal{P}(y = 1) \qquad B_7 \stackrel{def}{=} \mathcal{T}2\mathcal{P}(y = 2)$$
$$B_4 \stackrel{def}{=} \mathcal{T}2\mathcal{P}(x + y > 3) \qquad B_8 \stackrel{def}{=} \mathcal{T}2\mathcal{P}(x \geq 0).$$

---

7. In this case, if the SMT solver did not provide the original $\mathcal{T}$-lemmas when the feature of using mixed Boolean+theory-conflict clauses is active, then the latter feature should be disabled.

*The propositional formula* (10) *is unsatisfiable (fact (ii)), as demonstrated by the following resolution proof.*

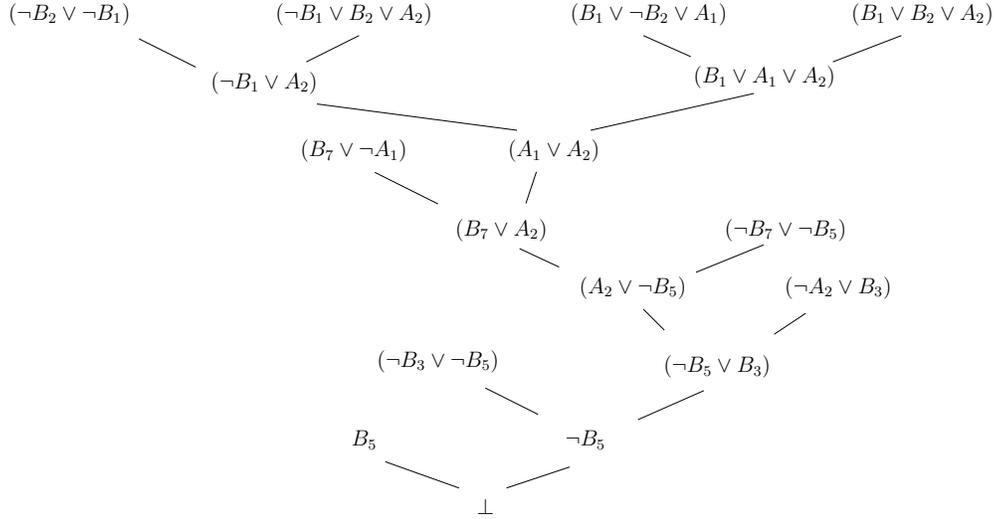

Fact (ii) holds also for those SMT tools which learn mixed Boolean+theory-clauses $F_1, \ldots, F_n$ (instead of $\mathcal{T}$-lemmas), obtained from the $\mathcal{T}$-lemmas $D_1, \ldots, D_n$ by backward traversal of the implication graph. In fact, in this case, $\mathcal{T}2\mathcal{P}(\varphi \wedge \bigwedge_{i=1}^{n} F_i) \models \bot$ holds. Since $\varphi \wedge \bigwedge_{i=1}^{n} D_i \models \bigwedge_{i=1}^{n} F_i$, because of the way the $F_i$'s are built, [8] (ii) holds.

Some SMT tools implement theory-propagation in a slightly different way (e.g. BARCE-LOGIC, Nieuwenhuis et al., 2006). If $l_1, \ldots, l_n \models_\mathcal{T} l$, instead of learning the $\mathcal{T}$-lemma $\neg l_1 \vee \ldots \vee \neg l_n \vee l$ and unit-propagating $l$ on it, they simply propagate the value of $l$, without learning any clause. Only if such propagation leads to a conflict later in the search, the theory-deduction clause is learned and used for conflict-analysis. The validity of fact (ii) is not affected by this optimization, because only the $\mathcal{T}$-lemmas used during conflict analysis are needed for it to hold (Nieuwenhuis et al., 2006).

Overall, in all variants of the on-line schema, the embedded DPLL engine builds –either explicitly or implicitly– a resolution refutation of the Boolean abstraction of the conjunction of the original clauses and the $\mathcal{T}$-lemmas returned by the $\mathcal{T}$-solver. Thus fact (ii) holds.

### 4.2 Extracting SMT Cores by Lifting Theory Lemmas

Facts (i) and (ii) discussed in §4.1 suggest a new approach to the generation of unsatisfiable cores for SMT. The main idea is that if the theory lemmas used during the SMT search are lifted into Boolean clauses, then the unsat core can be extracted by a purely propositional core extractor. Therefore, we call this technique the *Lemma-Lifting approach*.

The algorithm is presented in Figure 4. The procedure $\mathcal{T}$-`Unsat_Core` receives as input a set of clauses $\varphi =_{def} \{C_1, \ldots, C_n\}$ and it invokes on it a lazy $SMT(\mathcal{T})$ tool `Lazy_SMT_Solver`, which is instructed to *store* somewhere the $\mathcal{T}$-lemmas returned by the

---

[8]. Each clause $\mathcal{T}2\mathcal{P}(F_i)$ is obtained by resolving the clause $\mathcal{T}2\mathcal{P}(D_i)$ with clauses in $\mathcal{T}2\mathcal{P}(\varphi \wedge \bigwedge_{j=1}^{i-1} F_j)$, so that $\mathcal{T}2\mathcal{P}(\varphi \wedge \bigwedge_{j=1}^{i-1} F_j \wedge D_i) \models \mathcal{T}2\mathcal{P}(F_i)$. Thus, by induction, $\mathcal{T}2\mathcal{P}(\varphi \wedge \bigwedge_{i=1}^{n} D_i) \models \mathcal{T}2\mathcal{P}(\bigwedge_{i=1}^{n} F_i)$, so that $\varphi \wedge \bigwedge_{i=1}^{n} D_i \models \bigwedge_{i=1}^{n} F_i$.

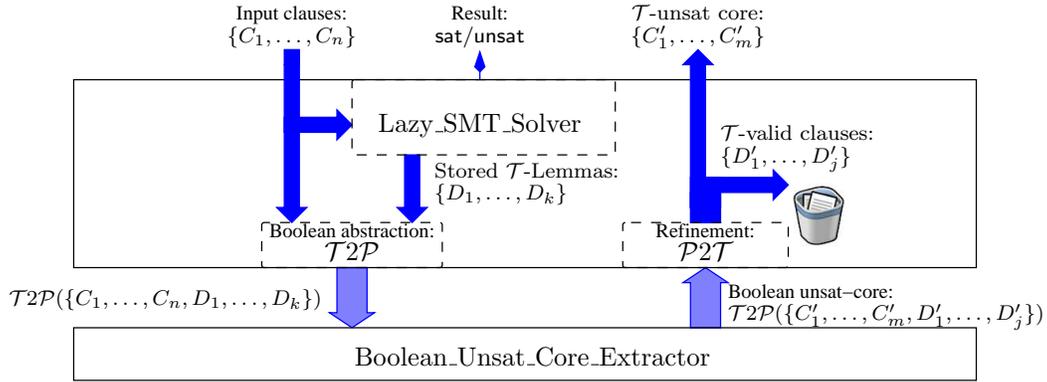

```
⟨SatValue,Clause_set⟩ T-Unsat_Core(Clause_set φ) {
    // φ is {C_1,...,C_n}
    if (Lazy_SMT_Solver(φ) == sat)
        then return ⟨sat,∅⟩;
    // D_1,...,D_k are the T-lemmas stored by Lazy_SMT_Solver
    ψ^p=Boolean_Unsat_Core_Extractor(T2P({C_1,...,C_n,D_1,...,D_k}));
    // ψ^p is T2P({C'_1,...,C'_m,D'_1,...,D'_j}));
    return ⟨unsat,{C'_1,...,C'_m}⟩;
}
```

Figure 4: Schema of the $\mathcal{T}$-Unsat_Core procedure: architecture (above) and algorithm (below).

$\mathcal{T}$-*solver*, namely $D_1,\ldots,D_k$. If `Lazy_SMT_Solver` returns sat, then the whole procedure returns sat. Otherwise, the Boolean abstraction of $\{C_1,\ldots,C_n, D_1,\ldots,D_k\}$, which is inconsistent because of (ii), is fed to an external tool `Boolean_Unsat_Core`, which is able to return the Boolean unsat core $\psi^p$ of the input. By construction, $\psi^p$ is the Boolean abstraction of a clause set $\{C'_1,\ldots,C'_m,D'_1,\ldots,D'_j\}$ s.t. $\{C'_1,\ldots,C'_m\} \subseteq \{C_1,\ldots,C_n\}$ and $\{D'_1,\ldots,D'_j\} \subseteq \{D_1,\ldots,D_k\}$. As $\psi^p$ is unsatisfiable, then $\{C'_1,\ldots,C'_m,D'_1,\ldots,D'_j\}$ is $\mathcal{T}$-unsatisfiable. By (i), the $\mathcal{T}$-valid clauses $D'_1,\ldots,D'_j$ have no role in the $\mathcal{T}$-unsatisfiability of $\{C'_1,\ldots,C'_m,D'_1,\ldots,D'_j\}$, so that they can be thrown away, and the procedure returns unsat and the $\mathcal{T}$-unsatisfiable core $\{C'_1,\ldots,C'_m\}$.

Notice that the resulting $\mathcal{T}$-unsatisfiable core is not guaranteed to be minimal, even if `Boolean_Unsat_Core` returns minimal Boolean unsatisfiable cores. In fact, it might be the case that $\{C'_1,\ldots,C'_m\}\backslash\{C'_i\}$ is $\mathcal{T}$-unsatisfiable for some $C'_i$ even though $\mathcal{T}2\mathcal{P}(\{C'_1,\ldots,C'_m\}\backslash\{C'_i\})$ is satisfiable, because all truth assignments $\mu^p$ satisfying the latter are such that $\mathcal{P}2\mathcal{T}(\mu^p)$ is $\mathcal{T}$-unsatisfiable.

**Example 5** *Consider the unsatisfiable SMT formula $\varphi$ on $\mathcal{LA}(\mathbb{Z})$:*

$$\begin{aligned}\varphi \equiv\ & ((x=0) \vee (x=1)) \wedge (\neg(x=0) \vee (x=1)) \wedge ((x=0) \vee \neg(x=1)) \wedge \\ & (\neg(x=0) \vee \neg(x=1))\end{aligned}$$

*and its propositional abstraction $\mathcal{T}2\mathcal{P}(\varphi)$:*

$$\mathcal{T}2\mathcal{P}(\varphi) \equiv (B_1 \vee B_2) \wedge (\neg B_1 \vee B_2) \wedge (B_1 \vee \neg B_2) \wedge (\neg B_1 \vee \neg B_2).$$

*Then, $\mathcal{T}2\mathcal{P}(\varphi)$ is a minimal Boolean unsatisfiable core of itself, but $\varphi$ is not a minimal core in $\mathcal{LA}(\mathbb{Z})$, since the last clause is valid in this theory, and hence it can be safely dropped.*

The procedure can be implemented very simply by modifying the SMT solver so that to store the $\mathcal{T}$-lemmas and by interfacing it with some state-of-the-art Boolean unsat core extractor used as an external black-box device. Moreover, if the SMT solver can provide the set of all $\mathcal{T}$-lemmas as output, then the whole procedure may reduce to a control device interfacing with both the SMT solver and the Boolean core extractor as black-box external devices.

**Remark 2** *Notice that here **storing** the $\mathcal{T}$-lemmas does not mean **learning** them, that is, the SMT solver is not required to add the $\mathcal{T}$-lemmas to the formula during the search. Instead, it is for instance sufficient to store them in some ad-hoc data structure, or even to dump them to a file. This causes no overhead to the Boolean search in the SMT solver, and imposes no constraint on the lazy strategy adopted (e.g., offline/online, permanent/temporary learning, usage of mixed Boolean+theory conflict clauses, etc.).*

**Example 6** *Once again, consider formula (1) of Example 1, and the corresponding formula (10) of Example 4, which is the Boolean abstraction of (1) and the $\mathcal{LA}(\mathbb{Z})$-lemmas (9) found by* MATHSAT *during search. In the Lemma-Lifting approach, (10) is given as input to an external Boolean unsat core device. The resulting propositional unsatisfiable core is:*

$$\begin{aligned}&\{(B_1 \vee \neg B_2 \vee A_1), (B_1 \vee B_2 \vee A_2), (\neg B_1 \vee B_2 \vee A_2), (\neg A_2 \vee B_3), B_5,\\ &(B_7 \vee \neg A_1), (\neg B_2 \vee \neg B_1), (\neg B_7 \vee \neg B_5), (\neg B_3 \vee \neg B_5)\},\end{aligned}$$

*which corresponds (via $\mathcal{P}2\mathcal{T}$) to:*

$$\begin{aligned}&\{((x=0) \vee \neg(x=1) \vee A_1), ((x=0) \vee (x=1) \vee A_2), (\neg(x=0) \vee (x=1) \vee A_2),\\ &(\neg A_2 \vee (y=1)), B_5, ((y=2) \vee \neg A_1),\\ &(\neg(x=1) \vee \neg(x=0)), (\neg(y=2) \vee \neg(y<0)), (\neg(y=1) \vee \neg(y<0))\}.\end{aligned}$$

*Since the last three clauses are included in the $\mathcal{LA}(\mathbb{Z})$-lemmas, and thus are $\mathcal{LA}(\mathbb{Z})$-valid, they are eliminated. The resulting core consists of only the first 6 clauses. In this case, the core turns out to be minimal, and is identical modulo reordering to that computed by* MATHSAT *with proof-tracing (see Example 1).*

As observed at the end of the previous section, our technique works also if the SMT tool learns mixed Boolean+theory clauses (provided that the original $\mathcal{T}$-lemmas are stored), or uses the lazy theory deduction of Nieuwenhuis et al. (2006). Moreover, it works also if $\mathcal{T}$-lemmas contain *new atoms* (i.e. atoms that do not appear in $\varphi$), as in the approaches of Flanagan et al. (2003), and Barrett, Nieuwenhuis, Oliveras, and Tinelli (2006), since both Facts (ii) and (i) hold also in that case.

As a side observation, we remark that the technique works also for the *per-constraint-encoding* eager SMT approach of Goel, Sajid, Zhou, Aziz, and Singhal (1998), and Strichman, Seshia, and Bryant (2002). In the eager SMT approach, the input $\mathcal{T}$-formula $\varphi$ is translated into an equi-satisfiable Boolean formula, and a SAT solver is used to check its satisfiability. With per-constraint-encoding of Goel et al. (1998) and Strichman et al. (2002), the resulting Boolean formula is the conjunction of the propositional abstraction $\varphi^p$ of $\varphi$ and a formula $\varphi^\mathcal{T}$ which is the propositional abstraction of the conjunction of some $\mathcal{T}$-valid clauses. Therefore, $\varphi^\mathcal{T}$ plays the role of the $\mathcal{T}$-lemmas of the lazy approach, and our approach still works. This idea falls out of the scope of this work, and is not expanded further.

### 4.3 Discussion

Despite its simplicity, the proposed approach is appealing for several reasons.

First, it is extremely simple to implement. The building of unsat cores is delegated to an external device, which is fully decoupled from the internal DPLL-based enumerator. Therefore, there is no need of implementing any internal unsat core constructor nor to modify the embedded Boolean device. Every possible external device can be interfaced in a plug-and-play manner by simply exchanging a couple of DIMACS files[9].

Second, the approach is fully compatible with optimizations carried out by the core extractor at the Boolean level: every original clause which the Boolean unsat core device is able to drop, is also dropped in the final formula. Notably, this involves also Boolean unsat-core techniques which could be very difficult to adapt to the SMT setting (and to implement within an SMT solver), such as the ones based on genetic algorithms (Zhang et al., 2006).

Third, it benefits for free from the research on propositional unsat-core extraction, since it is trivial to update: once some novel, more efficient or more effective Boolean unsat core device is available, it can be used in a plug-and-play way. This does not require modifying the DPLL engine embedded in the SMT solver.

One may remark that, in principle, if the number of $\mathcal{T}$-lemmas generated by the $\mathcal{T}$-solver were huge, the storing of all $\mathcal{T}$-lemmas might cause memory-exhaustion problems or the generation of Boolean formulas which are too big to be handled by the Boolean unsat-core extractor. In practice, however, this is not a real problem. In fact, even the hardest SMT formulas at the reach of current lazy SMT solvers rarely need generating more than $10^5$ $\mathcal{T}$-lemmas, whereas current Boolean unsat core extractors can handle formulas in the order of $10^6 - 10^7$ clauses. In fact, notice that the default choice in MathSAT is to learn all $\mathcal{T}$-lemmas permanently anyway, and we have never encountered problems due to this fact. Intuitively, unlike with plain SAT, in lazy SMT the computational effort is typically dominated by the search in the theory $\mathcal{T}$, so that the number of clauses that can be stored with a reasonable amount of memory, or which can be fed to a SAT solver, is typically much bigger than the number of calls to the $\mathcal{T}$-solver which can overall be accomplished within a reasonable amount of time.

---

9. DIMACS is a standard format for representing Boolean CNF formulas.

Like with the other SMT unsat-core techniques adopted by current SMT solvers, also with our novel approach the resulting $\mathcal{T}$-unsatisfiable core is not guaranteed to be minimal, even if `Boolean_Unsat_Core` returns minimal Boolean unsatisfiable cores. However, with the Lemma-Lifting technique it is possible to perform all the reductions that can be done by considering only the Boolean skeleton of the formula. Although this is in general not enough to guarantee minimality, it is still a very significant gain, as we shall show in the next section. Moreover, notice that it is also possible to obtain minimal UC's by iteratively calling one SMT core extractor, each time dropping one (or more) clause(s) from the current UC and checking for the $\mathcal{T}$-inconsistency. This minimization technique is orthogonal wrt. the SMT core-extractor adopted, and as such it is not investigated here.

## 5. Empirical Evaluation

We carried out an extensive experimental evaluation of the the Lemma-Lifting approach. We implemented the approach within the MathSAT (Bruttomesso et al., 2008) system. MathSAT has been extended with an interface for external Boolean unsatisfiable core extractors (UCE) to exchange Boolean formulas and relative cores in form of files in DIMACS format. (No modification was needed for the storage of $\mathcal{T}$-lemmas, because MathSAT already can learn permanently all of them.)

We have tried eight different external UCEs, namely Amuse (Oh et al., 2004), PicoSAT (Biere, 2008), Eureka (Dershowitz et al., 2006), MiniUnsat (van Maaren & Wieringa, 2008), MUP (Huang, 2005), Trimmer (Gershman et al., 2008), ZChaff (Zhang & Malik, 2003), and the tool proposed by Zhang et al. (2006) (called Genetic here). All these tools explicitly target core size reduction (or minimality), with the exception of PicoSAT, which was conceived for speeding up core generation, with no claims of minimality. In fact, PicoSAT turned out to be both the fastest and the least effective in reducing the size of the cores. For these reasons, we adopted it as our baseline choice, as it is the ideal starting point for evaluating the trade-off between efficiency (in execution time) and effectiveness (in core size reduction). Thus, we start evaluating our approach by using PicoSAT as external UCE (§5.1) and then we investigate the usage of more effective though more expensive UCE's (§ 5.2).

All the experiments have been performed on a subset of the SMT-LIB (Ranise & Tinelli, 2006) benchmarks. We used a total of 561 $\mathcal{T}$-unsatisfiable problems, taken from the QF_UF (126), QF_IDL (89), QF_RDL (91), QF_LIA (135) and QF_LRA (120) divisions, selected using the same criteria used in the annual SMT competition. In particular, the benchmarks are selected randomly from the available instances in the SMT-LIB, but giving a higher probability to real-world instances, as opposed to randomly generated or handcrafted ones. (See http://www.smtcomp.org/ for additional details.)

We used a preprocessor to convert the instances into CNF (when required), and in some cases we had to translate them from the SMT language to the native language of a particular SMT solver.[10]

---

10. In particular, CVC3 and Yices can compute unsatisfiable cores only if the problems are given in their own native format.

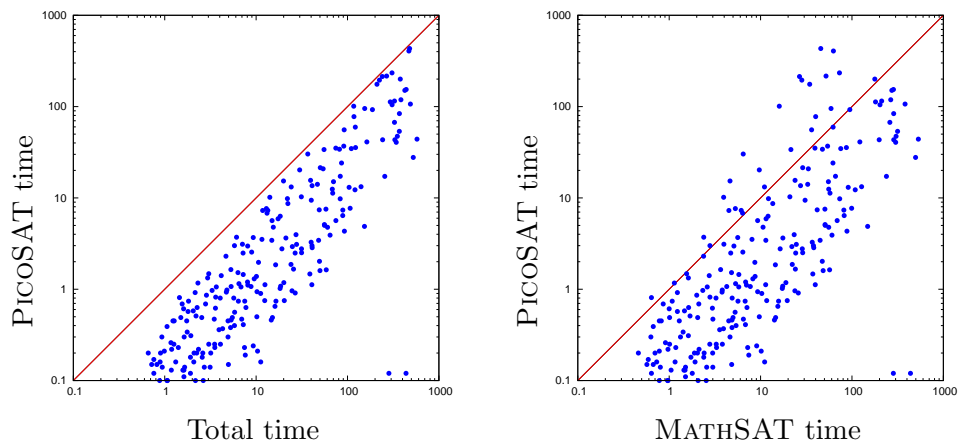

Figure 5: Overhead of PicoSAT wrt. the total execution time of MathSAT+PicoSAT (left) and wrt. the execution time of MathSAT (right).

All the tests were performed on 2.66 GHz Intel Xeon machines with 16 GB of RAM running Linux. For each tested instance (unless explicitly stated otherwise) the timeout was set to 600 seconds, and the memory limit to 2 GB. For all the Boolean UCEs, we have used the default configurations.

### 5.1 Costs and Effectiveness of Unsat-Core Extraction Using PicoSAT

The two scatter plots in Figure 5 give a first insight on the price that the Lemma-Lifting approach has to pay for running the external UCE. The plot on the left compares the execution time of PicoSAT with the total time of MathSAT+PicoSAT, whilst the plot on the right shows the comparison of the time of PicoSAT against that of MathSAT solving time only. From the two figures, it can be clearly seen that, except for few cases, the time required by PicoSAT is much lower or even negligible wrt. MathSAT solving time. Notice also that this price is payed only in the case of unsatisfiable benchmarks.

We now analyze our LL approach with respect to the size of the unsat cores returned. We compare the baseline implementation of our LL approach, MathSAT+PicoSAT, against MathSAT+ProofBasedUC (i.e. MathSAT with proof tracing), CVCLite (Barrett & Tinelli, 2007), [11] and Yices. [12] We have also performed a comparison with (the SMT version of) CAMUS (Liffiton & Sakallah, 2008), running it in "SingleMUS" mode (generate only one minimal UC, "CAMUS-one" hereafter). We also tried to run CAMUS in "AllMUS" mode (generate all minimal UC's), but we encountered some unexpected results (in some

---

[11]. We tried to use the newer CVC3, but we had some difficulties in the extraction of unsatisfiable cores with it. Therefore, we reverted to the older CVCLite for the experiments.
[12]. CVCLite version `20061231` and Yices version `1.0.19`.

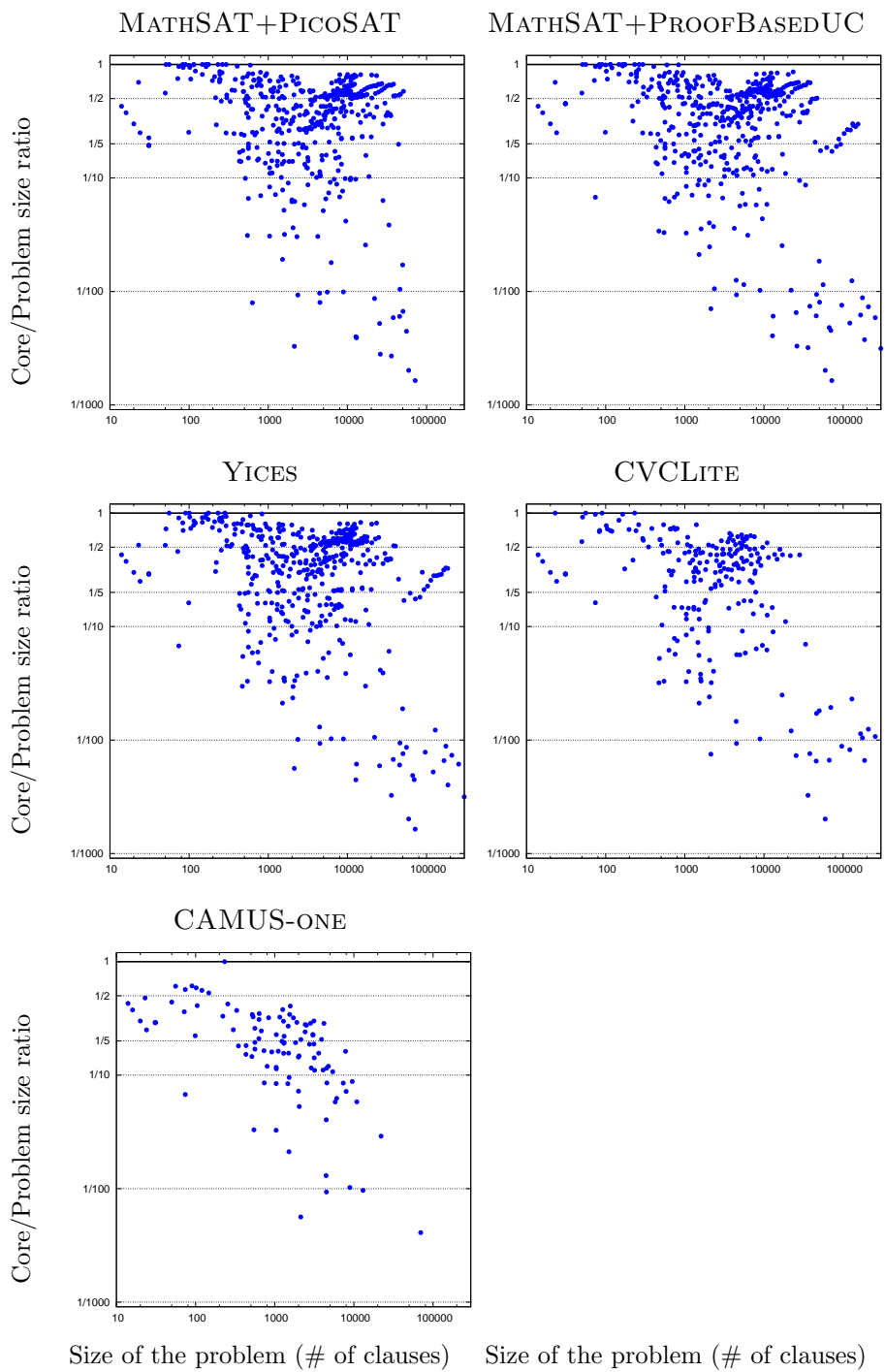

Figure 6: Ratio between the size of the original formula and that of the unsat core computed by the various solvers.

Cimatti, Griggio, & Sebastiani

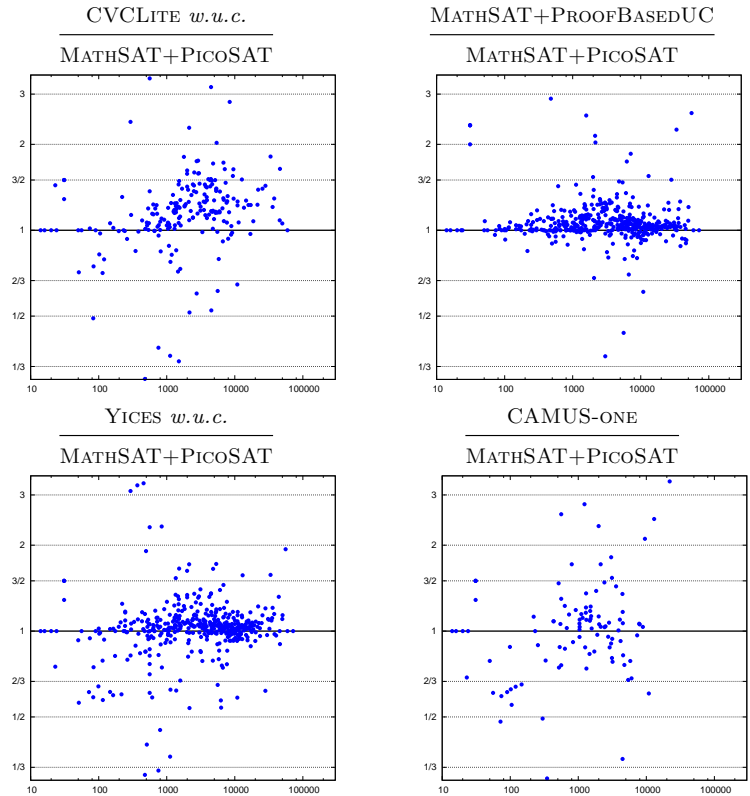

| core size ratio | $1^{st}$ quartile | median | mean | $3^{rd}$ quartile |
|---|---|---|---|---|
| $\dfrac{\text{CVCLite } w.u.c.}{\text{MathSAT+PicoSAT}}$ | 1.00 | 1.16 | 1.33 | 1.36 |
| $\dfrac{\text{MathSAT+ProofBasedUC}}{\text{MathSAT+PicoSAT}}$ | 1.00 | 1.03 | 1.09 | 1.10 |
| $\dfrac{\text{Yices } w.u.c.}{\text{MathSAT+PicoSAT}}$ | 0.97 | 1.03 | 1.08 | 1.09 |
| $\dfrac{\text{CAMUS-one}}{\text{MathSAT+PicoSAT}}$ | 0.88 | 1.02 | 1.32 | 1.18 |

Figure 7: Comparison of the size of the unsat cores computed by MathSAT+PicoSAT against those of CVCLite, MathSAT+ProofBasedUC, Yices with unsat cores and CAMUS-one, with statistics on unsat core ratios.
Points above the middle line and values greater than 1.00 mean better core quality for MathSAT+PicoSAT, and vice versa.

executions all the generated MUSes were larger than the unsat cores found by the other tools[13]), and so we had to exclude it from the experiments.

---

13. This is very surprising because, by definition, the output produced by CAMUS in "AllMUS" mode should should always contain UC's of minimum size, and thus smaller than those found by the other tools. Therefore, we have no explanation for such results, apart from conjecturing the presence of some bug in CAMUS, or some incorrect use from our side (although we followed the indications of the authors),

In order to allow CAMUS-ONE terminate for a significant amount of samples, we have run it with an increased timeout of 1800 seconds. Even so, CAMUS-ONE was able to produce one UC within the timeout only for 144 formulas out of 561. For the record, MATHSAT+PICOSAT, MATHSAT+PROOFBASEDUC, CVCLITE, and YICES solved within the timeout 474, 503, 253 and 494 problems out of 561 respectively.

Notice that we do not present any comparison in time between the different tools because it is not significant for determining the relative cost of unsat-core computation, since (i) for all the former four tools the time is completely dominated by the solving time, which varies a lot from solver to solver (even within MATHSAT, proof production requires setting ad-hoc options, which may result into significantly-different solving times since a different search space is explored); (ii) a comparison with CAMUS in terms of speed would not be fair, since the ultimate goal of CAMUS is to enumerate all mimimal UC's, and as such it first runs the very-expensive step of enumerating all MCS's (see §3.2).

Figure 6 shows the absolute reduction in size performed by the different solvers: the x-axis displays the size (number of clauses) of the problem, whilst the y-axis displays the ratio between the size of the unsat core and the size of the problem. For instance, a point with y value of 1/10 means that the unsatisfiability is due to only 10% of the problem clauses.

Figure 7(top) shows relative comparisons of the data of Figure 6. Each plot compares MATHSAT+PICOSAT with each of the other solvers. Such plots, which we shall call "core-ratio" plots, have the following meaning: the x-axis displays the size (number of clauses) of the problem, whilst the y-axis displays the ratio between the size of the unsat core computed by CVCLITE, MATHSAT+PROOFBASEDUC, YICES or CAMUS-ONE and that computed by MATHSAT+PICOSAT. For instance, a point with y value of 1/2 means that the unsat core computed by the current solver is half the size of that computed by MATHSAT+PICOSAT; values above 1 mean a smaller core for MATHSAT+PICOSAT. In core-ratio plots, we only consider the instances for which both solvers terminated successfully, since here we are only interested in the size of the cores computed, and not in the execution times. Figure 7(bottom) reports statistics about the ratios of the unsat core sizes computed by two different solvers.

A comment is in order. The results reported for CAMUS-ONE are quite surprising wrt. our expectations, since CAMUS-ONE is supposed to return a *minimal* UC, so that we would expect greater reductions in core sizes. This can be explained by the fact that the minimal UC produced by CAMUS-ONE is not necessarily *minimum*. In fact, we have manually verified for the samples with the biggest core-size ratio that the UC's returned by CAMUS-ONE are actually minimal, although significantly bigger than those returned by MATHSAT+PICOSAT.

Overall, the results presented show that, even when using as Boolean UCE PICOSAT, which is the least effective in reducing the *size* of the cores, the effectiveness of the baseline version of our LL approach is slightly better than those of the other tools.

---

or the activation by default of some of the incomplete heuristics CAMUS can use in order to cope with the combinatorial explosion in the number of MCS's UC's generated (see §3.2.)

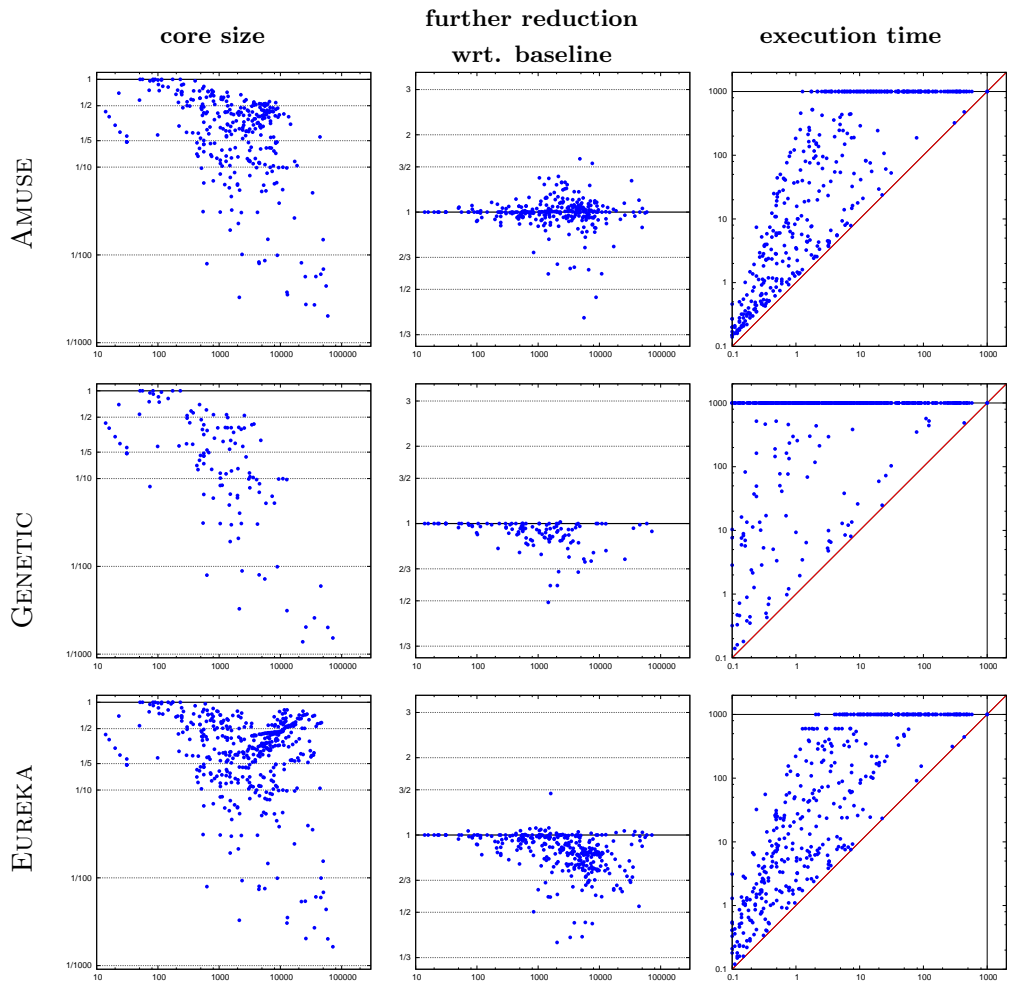

Figure 8: Comparison of the core sizes (left), core ratios (middle) and run times (right) using different propositional unsat core extractors. In the core-ratio plots ($2^{nd}$ column), the X-axis represents the size of the problem, and the Y-axis represents the ratio between the size of the cores computed by the two systems: a point above the middle line means better quality for the baseline system. In the scatter plots ($3^{rd}$ column), the baseline system (MathSAT+PicoSAT) is always on the X-axis.

## 5.2 Impact on Costs and Effectiveness Using Different Boolean Unsat Core Extractors

In this second part of our experimental evaluation we compare the results obtained using different UCE's in terms of costs and effectiveness in reducing the size of the core. We show that, depending on the UCE used, it is possible to reduce significantly the size of cores, and to trade core quality for speed of execution (and vice versa), with no implementation effort. We compare our baseline configuration MathSAT+PicoSAT, against six other configurations, each calling a different propositional UCE.

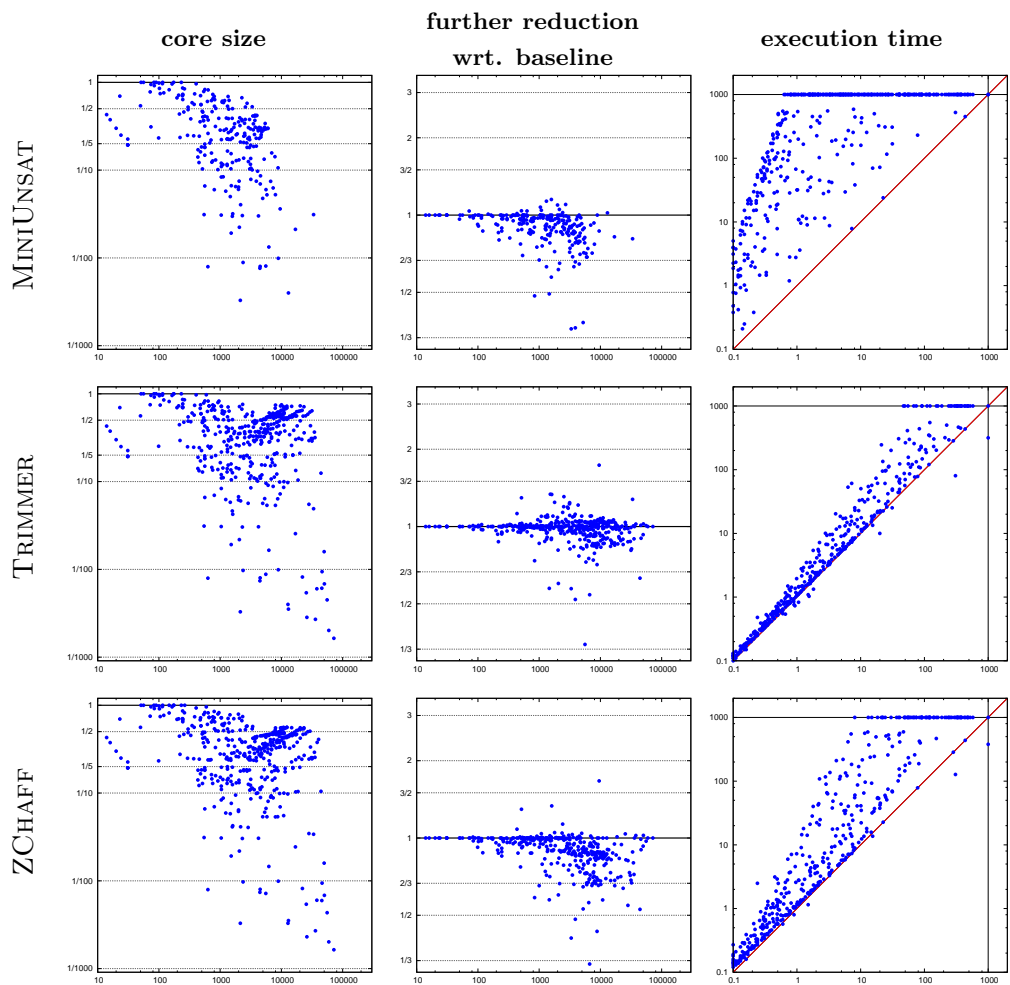

Figure 9: Comparison of the core sizes (left), core ratios (middle) and run times (right) using different propositional unsat core extractors (continued).

The results are collected in Figures 8-9. The first column shows the absolute reduction in size performed by each tool (as in Figure 6). The second column shows core-ratio plots comparing each configuration against the baseline one using PicoSAT (as in Figure 7, with points below 1.00 meaning a better performance of the current configuration). Finally, the scatter plots in the third column compare the execution times (with PicoSAT always on the X-axis). We evaluated the six configurations which use, respectively, Amuse (Oh et al., 2004), Genetic (Zhang et al., 2006), Eureka (Dershowitz et al., 2006), MiniUnsat (van Maaren & Wieringa, 2008), Trimmer (Gershman et al., 2008), and ZChaff (Zhang & Malik, 2003), against the baseline configuration, using PicoSAT. We also compared with MUP (Huang, 2005), but we had to stop the experiments because of memory exhaustion problems. Looking at the second column, we notice that Eureka, followed by MiniUnsat and ZChaff, seems to be the most effective in reducing the size of the final unsat cores, up to 1/3 the size of those obtained with plain PicoSAT. Looking at the third column, we notice that with Genetic, Amuse, MiniUnsat and ZChaff, and in part with Eureka, efficiency degrades drastically, and many problems cannot be solved within the timeout. With Trimmer the performance gap is not that dramatic, but still up to an order magnitude slower than the baseline version.

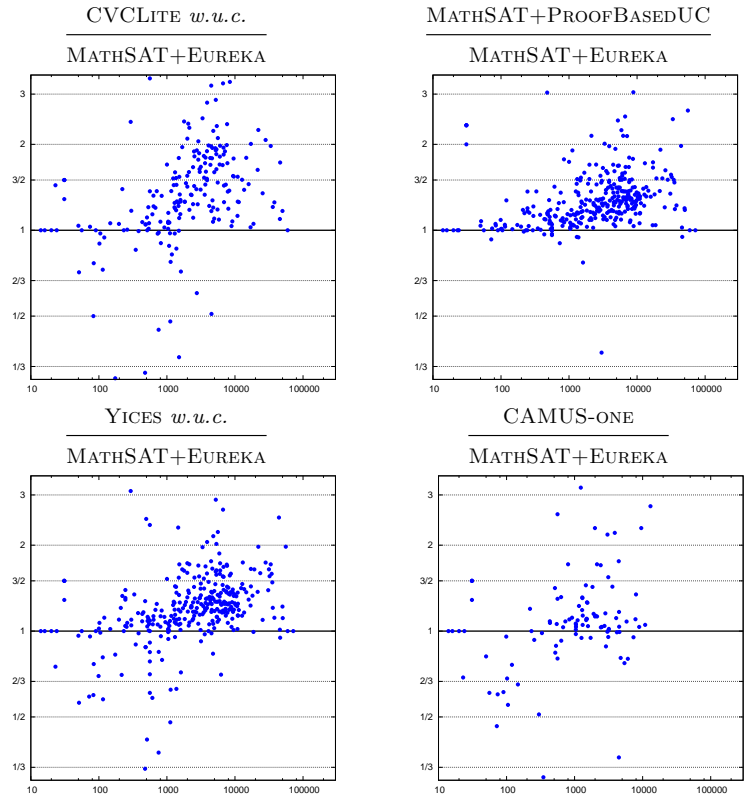

| core size ratio | $1^{st}$ quartile | median | mean | $3^{rd}$ quartile |
|---|---|---|---|---|
| $\dfrac{\text{CVCLite } w.u.c.}{\text{MathSAT+Eureka}}$ | 1.03 | 1.32 | 1.55 | 1.73 |
| $\dfrac{\text{MathSAT+ProofBasedUC}}{\text{MathSAT+Eureka}}$ | 1.03 | 1.17 | 1.27 | 1.35 |
| $\dfrac{\text{Yices } w.u.c.}{\text{MathSAT+Eureka}}$ | 1.00 | 1.16 | 1.28 | 1.34 |
| $\dfrac{\text{CAMUS-one}}{\text{MathSAT+Eureka}}$ | 0.98 | 1.05 | 1.41 | 1.26 |

Figure 10: Ratios of the unsat-core sizes computed by MathSAT+Eureka against those of CVCLite, MathSAT+ProofBasedUC, Yices and CAMUS-one. Points above the middle line and values greater than 1.00 mean better core quality for MathSAT+Eureka, and vice versa.

Finally, in Figure 10 we compare the effectiveness of MathSAT+Eureka, the most effective extractor in Figures 8-9, directly with that of the other three solvers, CVCLite, MathSAT+ProofBasedUC and Yices, and with that of CAMUS. (Also compare the results with those in Figure 7.) The gain in core reduction wrt. previous state-of-the-art SMT core-reduction techniques is evident.

It is important to notice that, due to our limited know-how, we used the Boolean UCE's in their default configurations. Therefore, we believe that even better results, in terms of both effectiveness and efficiency, could be obtained by means of a more accurate tuning of the parameters of the core extractors.

As a side remark, we notice that the results in Figures 8-9 have produced as a byproduct an insightful evaluation of the main Boolean unsat-core-generation tools currently available. To this extent, we notice that the performances of MUP (Huang, 2005) and Genetic (Zhang et al., 2006) seem rather poor; PicoSAT (Biere, 2008) is definitely the fastest tool, though the least effective in reducing the size of the final core; on the opposite side, Eureka (Dershowitz et al., 2006) is the most effective in this task, but pays a fee in terms of CPU time; Trimmer (Gershman et al., 2008) represents a good compromise between effectiveness and efficiency.

## 6. Conclusions

We have presented a novel approach to generating small unsatisfiable cores in SMT, that computes them a posteriori, relying on an external propositional unsat core extractor. The technique is very simple in concept, and straightforward to implement and update. Moreover, it benefits for free of all the advancements in propositional unsat core computation. Our experimental results have shown that, by using different core extractors, it is possible to reduce significantly the size of cores and to trade core quality for speed of execution (and vice versa), with no implementation effort.

As a byproduct, we have also produced an insightful evaluation of the main Boolean unsat-core-generation tools currently available.

### Acknowledgments

We wish to thank Mark Liffiton for his help with the CAMUS tool. We also thank the anonymous referees for their helpful suggestions.

A. Griggio is supported in part by the European Community's FP7/2007-2013 under grant agreement Marie Curie FP7 - PCOFUND-GA-2008-226070 "progetto Trentino", project Adaptation.

R. Sebastiani is supported in part by SRC under GRC Custom Research Project 2009-TJ-1880 WOLFLING.